\documentclass[conference,a4paper,flushend]{iaria}

\usepackage{amsmath,amssymb,amsfonts}
\usepackage{algorithmic}
\usepackage{graphicx}
\usepackage{textcomp}
\usepackage{xcolor}
\def\BibTeX{{\rm B\kern-.05em{\sc i\kern-.025em b}\kern-.08em
    T\kern-.1667em\lower.7ex\hbox{E}\kern-.125emX}}
\usepackage[style=ieee,backend=biber]{biblatex}
\usepackage{orcidlink} %https://ctan.org/pkg/orcidlink
\usepackage[T1]{fontenc}
\usepackage{paralist} % for compactitem environment
\usepackage{makecell}
\usepackage{booktabs}
    
\begin{document}

% \title{Noisy Networks, Nosy Neighbors: A Simple Attack Model against Neighboring Traffic}
\title{Noisy Networks, Nosy Neighbors: Simple Privacy Attacks Against Residential Wireless Traffic}

% Original: Noisy networks, nosy neighbors: Inferring privacy invasive information from encrypted wireless traffic

% Vorschläge:
% Noisy Networks, Nosy Neighbors: A Simple Attack Model against Neighboring Apartments
% Noisy Networks, Nosy Neighbors: A Simple Attack Model against Nearby Wireless Traffic in Neighboring Apartments
% Noisy Networks, Nosy Neighbors: A Simple Attack Model against Residential Wireless Traffic
% Noisy Networks, Nosy Neighbors: A Simple Attack Model against Wireless Traffic in Neighboring Apartments
% Noisy Networks, Nosy Neighbors: A Simple Privacy Attack against Residential Wireless Traffic

% AI Slop:
% Noisy Networks, Nosy Neighbors: Uncovering Hidden Patterns in Encrypted Wi‑Fi Streams
% Noisy Networks, Nosy Neighbors: Leveraging Ambient RF Emissions for Apartment‑Level Surveillance
% Noisy Networks, Nosy Neighbors: Side‑Channel Analysis of Home‑Router Handshakes
% Noisy Networks, Nosy Neighbors: Inferring Occupancy and Device Types from Signal Noise
% Noisy Networks, Nosy Neighbors: A Simple Threat Model for Residential Wi‑Fi Eavesdropping
% Noisy Networks, Nosy Neighbors: Quantifying Privacy Leakage in Dense Urban WLAN Deployments
% Noisy Networks, Nosy Neighbors: Passive Reconnaissance of Encrypted Traffic Across Shared Walls
% Noisy Networks, Nosy Neighbors: Exploiting Temporal Correlations in Neighboring Wireless Channels
% Noisy Networks, Nosy Neighbors: Detecting Covert Communications via Ambient Spectrum Analysis
% Noisy Networks, Nosy Neighbors: Assessing the Feasibility of Cross‑Apartment Traffic Fingerprinting 

\author{
\IEEEauthorblockN{%
Arne Roszeitis\IEEEauthorrefmark{1},
Bartosz Burgiel\IEEEauthorrefmark{2},
Victor J\"uttner\IEEEauthorrefmark{1},
Erik Buchmann\IEEEauthorrefmark{1},
}
\IEEEauthorblockA{\IEEEauthorrefmark{1}Center for Scalable Data Analytics and Artificial Intelligence (ScaDS.AI) Dresden/Leipzig, Leipzig University, Leipzig, Germany \\
Email: \{arne.roszeitis, victor.juettner, erik.buchmann\}@uni-leipzig.de}
\IEEEauthorblockA{\IEEEauthorrefmark{2}DigiFors GmbH, Leipzig, Germany \\
Email: bartosz.burgiel@digifors.de}
}

\maketitle

\begin{abstract}
Smart devices, such as light bulbs, TVs, fridges, etc., equipped with computing capabilities and wireless communication, are part of everyday life in many households. Previous work has already shown that a passive eavesdropper can derive private information, household routines, etc., from the network traffic of smart devices.
% Passive network eavesdroppers can infer sensitive household routines from encrypted smart-home traffic using only wireless metadata. 
However, existing attacks rely on capable adversaries with specialized machine learning expertise, labeled training data and reference devices, leaving it unclear how vulnerable ordinary households are to less sophisticated attackers. In this paper, we investigate the extent to which a ``casual attacker'' with straightforward IT skills and no specialized cybersecurity or ML tooling can reproduce such privacy attacks. Operating from an adjacent room in a real-world apartment building, we constrain our adversary to use only three off-the-shelf Raspberry Pis, Wireshark, and basic Python scripts. Through a three-week study, we demonstrate that this casual attacker can manually identify devices, recognize user states, track smartphone movements through walls via RSSI triangulation, and successfully extract detailed daily routines, including sleep patterns of guests. Our findings show that smart-home privacy leakage is a threat even from low-resourced, straightforward adversaries, e.g., neighbors.
\end{abstract}


\section{Introduction}
\label{sec:intro}

% Motivation
Smart devices like light bulbs, TVs, and plugs have become a routine part of many households. They continuously communicate via wireless protocols like WiFi, Bluetooth Low Energy (BLE), or ZigBee for control, updates and cloud connectivity. 
Although the payload of the wireless communication is encrypted, the plain-text metadata of the protocols~\cite{apthorpe2017spying,apthorpe2016poster} allow privacy attacks. Using timing, packet sizes and radio characteristics, a passive attacker in an adjacent apartment can reveal which devices are present, when and where they are active, and ultimately the daily routines of the residents. % All without decrypting a single packet. 

% Problem description
Prior work has shown that this threat is real and technically well-understood. Approaches, such as Peek-a-Boo~\cite{acar2020peekaboo}, IoTBeholder~\cite{zou2023iotbeholder}, WiFinger~\cite{Li2025WiFingerFN} and PingPong~\cite{trimananda2019pingpong} demonstrate that encrypted smart-home traffic is sufficient to classify device types, infer device states, localize devices and perform Human Activity Recognition (HAR) on residents. However, all of these approaches assume a capable adversary: they require expert knowledge of machine learning pipelines, access to labeled training data or reference devices, and have been evaluated in carefully controlled smart-home testbeds.

% Research question
What remains unclear is whether such attacks are also within reach of a casual attacker. We define a casual attacker as someone with a straightforward IT background, such as an undergraduate computer science student or a tech-savvy hobbyist. This attacker understands fundamental networking concepts and can write simple data-processing scripts, but lacks specialized tooling for cybersecurity or machine learning tasks. They operate with commodity hardware and have no prior access to the target's floor plan or devices to collect training data. A motivated but technically modest attacker of this profile, such as a curious neighbor, presents a realistic threat. Therefore, we investigate the following research question: 

\textit{Can a casual attacker reproduce smart-home privacy attacks in a real-world apartment?}

% Method / pipeline
To investigate this, we conduct a three-week study in a private apartment equipped with typical smart devices, with the consent of the residents. For this experiment, we assume the role of the casual attacker, eavesdropping on the apartment. To capture the WiFi and BLE traffic through the wall, the attacker is allowed to use only three Raspberry Pis. Furthermore, the subsequent data analysis is strictly limited to Wireshark and basic Python scripts for processing network packets and visualization. Using only these tools we attempt to achieve four privacy-invasive goals via a manual, heuristic pipeline: 
(1)~device identification by combining publicly available MAC OUI (Organizationally Unique Identifier) lookups and BLE advertisements; 
(2)~device state recognition (e.g., TV on/off) by observing basic traffic volume thresholds; 
(3)~coarse indoor localization via basic RSSI triangulation from the three Raspberry Pis; and 
(4)~behavior inference and guest detection by applying typical behavior patterns to plotted network activity over time.

% Contributions
We make two main contributions:
\begin{compactitem}
    \item We demonstrate that a casual attacker with limited resources and expertise can reproduce smart-home privacy attacks in an uncontrolled, real-world setting.
    \item We empirically detail this threat, showing the attacker can identify devices, recognize user states, track smartphone movements through walls, and extract daily routines including the presence and sleep schedules of guests.
\end{compactitem}
Our findings demonstrate that smart-home privacy leakage is not limited to powerful adversaries with specialized equipment. This emphasizes the need for privacy mechanisms that protect against close-by, passive, and persistent eavesdroppers of wireless metadata~\cite{Wang2024ISS,apthorpe2017spying}. More detailed statistics and discussions can be found on a thesis~\cite{bartosz2025}.

Paper structure: Section~\ref{sec:related} reviews related work. Section~\ref{sec:method} outlines our research approach. The results of our experiments are shown in Section~\ref{sec:results} and then discussed in Section~\ref{sec:discussion}. Finally, Section~\ref{sec:conclusion} concludes the paper.

\section{Related Work}
\label{sec:related}

The related work provides state-of-the-art background as base for our attack pipeline and discusses existing smart-home traffic analysis and HAR attacks.

\subsection{Background:}

\paragraph*{Traffic analysis and device discovery.}
Encrypted smart-home traffic can be analyzed via metadata, such as IP/TCP headers, DNS queries and throughput. Apthorpe et al.\ show that an ISP-level observer, seeing only the gateway IP address, can infer device types by splitting traffic into substreams and inspecting DNS and rate information \cite{apthorpe2016poster}. Device discovery can also exploit link-layer identifiers: OUIs in MAC addresses reveal manufacturers, and probe requests expose preferred Service Set Identifiers (SSIDs).

\paragraph*{RSSI-based localization.}
Received Signal Strength Indicators (RSSI) are widely used for indoor localization of wireless devices. RADAR demonstrates in-building RF-based user tracking using multiple WiFi receivers and signal strength \cite{bahl2000RADAR}. Pérez Iglesias et al.\ apply Bluetooth RSSI fingerprinting to localize persons indoors with at least three receivers \cite{Iglesias2012IndoorPL}, and Mane et al.\ refine RSSI-based distance estimation in WiFi IoT settings using machine learning \cite{mane2026rssi_distance_ai}. Precision depends on power, range and environment, but high accuracy is achievable in favourable conditions.

\subsection{Smart-Home HAR and Traffic-Based Attacks}

Passive network adversaries sniffing traffic from IoT devices are a known threat \cite{apthorpe2017spying}. Acar et al.\ propose \textbf{Peek-a-Boo}, a multi-level attack that uses WiFi, ZigBee and BLE traffic to first identify devices, then infer their states and finally perform HAR on user activities \cite{acar2020peekaboo}. Their evaluation is conducted in a controlled smart-home testbed and assumes access to similar devices for training and prior knowledge of the device set. 

Zou et al.\ present \textbf{IoTBeholder}, which applies machine-learning pipelines to smart-home traffic and sensor data to build HAR models and predict user behaviour \cite{zou2023iotbeholder}. IoTBeholder integrates HAR deeply into the attack flow and targets habitual behaviours over time, but assumes expert-level data mining knowledge and carefully curated training datasets. 

Trimananda et al.\ introduce \textbf{PingPong}, which learns packet-level signatures for smart-home device events using neural networks trained on labeled traces under WAN and WiFi sniffer threat models \cite{trimananda2019pingpong}. PingPong assumes a known device set and requires labeled reference traces for training.

Li et al.\ propose \textbf{WiFinger}, which aims to infer IoT device states from WiFi traffic and uses packet-level sequence matching to remain robust under packet loss and network noise \cite{Li2025WiFingerFN}. WiFinger operates under a WiFi sniffer threat model similar to that used above and still requires prior knowledge of the devices and access to similar devices for training machine-learning models. 

Beyond general device and state inference, Wang et al.\ examine two voice-controlled devices and collect their encrypted traffic to infer the answers to voice prompts \cite{Wang2020FingerprintingEV}. They use CNN, LSTM and stacked autoencoder models in an ensemble, assuming access to the local network and substantial machine-learning expertise. Collectively, these works show that encrypted smart-home traffic can be exploited for device-level and state inference, HAR and even content-related inference when attackers have access to suitable training data and controlled setups. Table~\ref{tab:related-comparison} summarizes the attacker expertise in related work compared to our setting.

\begin{table}[t]
\centering
\caption{related work on smart home privacy attacks and har}
\label{tab:related-comparison}
\begin{tabular}{p{2.3cm}p{5.5cm}}
\toprule
\textbf{Work} & \textbf{Attacker knowledge / resources} \\
\midrule
Peek-a-Boo \cite{acar2020peekaboo} & Similar devices, labeled traces, multi-protocol capture, ML expertise \\
IoTBeholder \cite{zou2023iotbeholder} & Expert ML pipeline, large training datasets \\
WiFinger \cite{Li2025WiFingerFN} & Prior device knowledge, access to similar devices, ML models \\
PingPong \cite{trimananda2019pingpong} & Trained neural networks, reference traces, known device set \\
Voice fingerp.~\cite{Wang2020FingerprintingEV} & Local network access, deep learning models \\
This work & Basic Python scripts, 3 Raspberry Pis \\
\bottomrule
\end{tabular}
\end{table}

\subsection{Research Gap}

Existing work demonstrates that encrypted smart-home traffic can be exploited for device-, state- and activity-level inference, but typically assumes skilled adversaries, labeled training data and carefully controlled testbeds. Our study instead evaluates how far similar inferences are possible for a casual attacker using commodity hardware and simple scripts in a realistic household.

\section{Methodology}
\label{sec:method}
In this section, we present our four-stage inference pipeline for the analysis, describe the experimental setup including the smart home environment and the monitoring station and detail the data collection procedure and pre-processing steps.

\subsection{The Inference Pipeline}
The inference pipeline is designed so that the casual attacker progressively builds knowledge about the smart home and its residents in four sequential steps:
\begin{enumerate}
    \item \textbf{Device Identification:} First, we search for and identify network devices by analyzing MAC OUIs, BLE advertisements, and unencrypted setup traffic.
    
    \item \textbf{Device State Recognition:} To determine each devices operational state (active, idle, or off), we monitor packet volumes over time and establish traffic thresholds.
    
    \item \textbf{Device Localization:} We then aim to map physical locations. Using RSSI trilateration from three distributed sensors, we approximate the apartment's layout and track mobile devices through walls.
    
    \item \textbf{Human Activity Recognition:} Finally, we synthesize this spatial and temporal data. By correlating smartphone movements with appliance states, we deduce sensitive daily routines like work schedules, sleep cycles, and guest presence.
\end{enumerate}

\subsection{Experimental Setup}
\begin{figure}[h!]
    \centering
    \includegraphics[width=.6\linewidth]{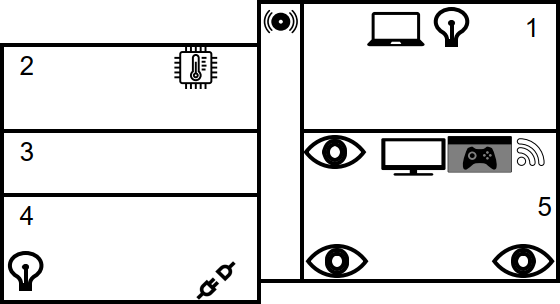}
    \caption{Floor plan of the experimental apartment: Office (1), Kitchen (2), Bath (3), Bedroom (4), and Living room (5).}

    \label{fig:appartment}
\end{figure}

The experiment was conducted in a five-room apartment equipped with a network of 10 to 12 wireless devices. Figure~\ref{fig:appartment} illustrates the floor plan, while Table~\ref{tab:devices} lists the deployed devices and specifies their exact room assignments. This device mix reflects a modern household, consisting of automated smart home appliances, stationary multimedia and work devices, mobile personal smartphones, and a central WiFi router.

\begin{table}[h!]
\centering
\caption{Smart Devices for our Study}
\begin{tabular}{p{1cm} p{3.4cm} p{1.2cm} p{0.5cm}}
\toprule
\textbf{Mac} & \textbf{Device} &  \textbf{Conn.} & \textbf{Room} \\ 
\midrule
d8:f1& Tuya light bulb BKL1259 &WiFi &1 \\
08:b6& Shelly Plus HT & WiFi/BT & 2\\
6c:5a& Tapo Bulb E225& WiFi&4\\
54:af& Tapo Smart Plug P100& WiFi/BT &4\\
8c:f6& Shelly Motion 2& WiFi &hall\\
9c:fc& Laptop (Intel) & WiFi/BT & 1\\
24:2f& TP-Link router& WiFi & 5\\
20:28& LG Smart TV UQ75009LF & WiFi/BT & 5\\
60:1a& Nintendo Switch 2 &WiFi &5\\
a4:45& Redmi Note 8 Pro & WiFi/BT & mobile\\
ae:90& Smartphone Guest 1 & WiFi/BT & mobile\\
e2:e2& Smartphone Guest 2 & WiFi/BT & mobile\\
\bottomrule
\end{tabular}
\label{tab:devices}
\end{table}

To mimic the casual attacker operating from an adjacent room, we established a monitoring setup in the living room. Three Raspberry Pi 4 Model B units (4 GB RAM, PI OS Lite) equipped with TP-Link TL-WN722N WiFi antennas were placed in the corners of the room to enable RSSI triangulation. 
% Corner of the living room! Symbolized by eyes in Figure 1
The adversary was restricted to standard software tools, specifically Wireshark~\cite{wireshark}, bluepy3~\cite{bluepy}, and tcpdump~\cite{tcpdump}. These tools were controlled via simple Python scripts to capture WiFi packets, BLE data, and RSSI readings without attempting to decrypt the payloads.

\subsection{Data Collection and Pre-Processing}
Data collection spanned a three-week period to capture a comprehensive picture of the resident's daily routines. This time frame included the initial factory-reset and installation of all smart devices, regular daily activities (e.g., leaving for university, leisure time at home), and a controlled evening visit by two guests carrying personal smartphones.

For pre-processing, the captured network traffic was categorized into three distinct datasets: general traffic data, signal strength vectors (RSSI), and probe request packets. To simulate the limitations of a casual adversary and to save storage space, all encrypted payloads were stripped from the traffic data, retaining only the payload length. Furthermore, both the traffic data and the RSSI values were aggregated into one-second intervals. The sanitized data was then visualized and analyzed using common Python libraries to execute the four objectives of our inference pipeline.

\section{Study Results}
\label{sec:results}
Following our inference pipeline, we demonstrate how a casual adversary can gradually reconstruct a detailed picture of the residents of the apartment, their devices and their routines.

\subsection{Device Identification}
The first step is to isolate the target network, inventory the devices, and categorize their fundamental behavior.

\paragraph{Passive Fingerprinting and Attribution}
By filtering network traffic based on signal strength, the adversary identifies frames belonging to the target apartment and thereby discovers its SSID. From there, all unique MAC addresses are extracted. Using simple Python scripts for OUI lookups~\cite{macaddressio,macvendors}, the attacker successfully attributes most smart appliances to their manufacturers (e.g., Tuya, TP-Link). Furthermore, captured BLE advertisements directly leak exact device models and names (e.g., "Complete Local Name: ShellyPlusHT-08B6"). Some smartphones and laptops utilize randomized MAC addresses that prevent direct OUI attribution, their presence is deduced from their complex, high-volume traffic.

\paragraph{Device Installation Procedure}
% Is the part that new devices are installed added clear enough in method/results?
The attacker gains a significant advantage when new devices are added to the network. During the coupling process, several smart appliances establish temporary, unencrypted WiFi access points. By eavesdropping on this setup traffic, the adversary extracts explicit device names (e.g., \textit{Tapo E225}), firmware versions, and embedded web server traffic. Most alarmingly, the companion apps transmitted the household's primary WiFi SSID and password in plain text. Concurrently, the resident's connected smartphone leaked mDNS service names (e.g., Google, Spotify), immediately revealing the resident's digital ecosystem.

\begin{table}[h!]
\centering
\caption{Attacker's knowledge after device identification.}
\label{tab:inferred_inventory}
\begin{tabular}{@{}llcc@{}}
\toprule
\textbf{MAC} & \textbf{Inferred Device (Source)} & \textbf{Profile} & \textbf{Mobility} \\
\midrule
d8:f1 & Espressif Smart Device \textit{(MAC)} & Auton. & Static \\
08:b6 & Shelly Plus HT \textit{(BLE)} & Auton. & Static \\
6c:5a & Tapo Bulb E225 \textit{(Installation)} & Auton. & Static \\
54:af & TP-Link Smart Device \textit{(MAC)} & Auton. & Static \\
8c:f6 & Shelly Motion 2 \textit{(Installation)} & Auton. & Static \\
9c:fc & Intel Multimedia \textit{(MAC)} & Inter. & Static \\
24:2f & TP-Link Router \textit{(MAC)} & Auton. & Static \\
20:28 & LG TV UQ75009LF \textit{(BLE)} & Inter. & Static \\
60:1a & Nintendo Switch \textit{(MAC)} & Inter. & Static \\
a4:45 & Redmi Note 8 Pro \textit{(Installation)} & Inter. & Mobile \\
\bottomrule
\multicolumn{4}{l}{\footnotesize \textit{Note: Two guest smartphones excluded from baseline.}}
\end{tabular}
\end{table}

\paragraph{Inferred Devices}
As summarized in Table~\ref{tab:inferred_inventory}, the adversary successfully profiles the household's 10 baseline resident devices (the two guest smartphones are absent at this stage). Crucially, the table reflects the attacker's exact state of knowledge: while some devices leaked exact model names via setup traffic or BLE, others were inferred broadly via OUI manufacturer lookups (e.g., inferring a gaming console from a ``Nintendo'' MAC prefix). Despite varying levels of identification precision, mapping the traffic profiles (interactive vs. autonomous) and mobility states (static vs. mobile) provides the necessary foundation to track device states and physical locations in the subsequent phases.

\subsection{Device State Recognition}
By plotting packet counts over time, the attacker establishes simple heuristic thresholds to classify the operational state of a device as \textit{active}, \textit{idle}, or \textit{off}. 

% \paragraph{Establishing Operational Baselines}
%To prepare for state recognition and localization, the attacker analyzes the baseline behavior of the identified devices. 
% To distinguish different kinds of devices, the attacker analyzes the baseline behavior of the monitored MAC addresses. 

\begin{figure}[h!]
\includegraphics[width=1.05\linewidth]{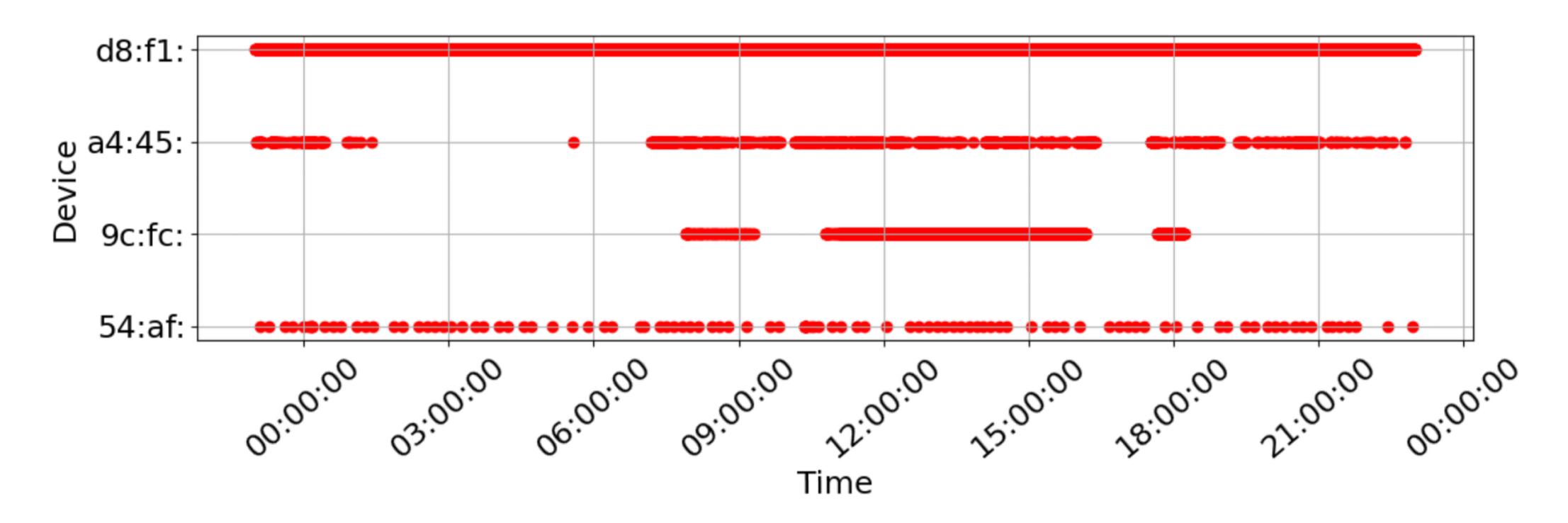}
\caption{Network activity over one day.}
\label{fig:device_discovery}
\end{figure}

As illustrated in Figure~\ref{fig:device_discovery}, observing network activity over 24 hours allows the attacker to classify devices into two distinct operational profiles based on their traffic signatures. First, \textit{autonomous} devices, such as the Tuya light bulb (d8:f1) and the Tapo smart plug (54:af) exhibit continuous, periodic background traffic regardless of user presence. Second, \textit{interactive} devices, such as the smartphone (a4:45) and the laptop (9c:fc) display dormant periods punctuated by sharp traffic spikes that directly correlate with active human usage. Additionally, by analyzing the variance in RSSI readings, the attacker distinguishes between \textit{stationary} appliances (which exhibit minor signal fluctuations) and \textit{mobile} devices. 
The third and fourth column of Table~\ref{tab:inferred_inventory} shows the profiles and mobility classes of the devices. 

% As previously identified, \textit{autonomous} smart home appliances (e.g., the Tuya bulb and Shelly sensors) exhibit continuous, periodic traffic that rarely changes significantly, making state inference in our experimental setup unreliable. 
% This is due to our simple setup though! In related work it has been shown that this is possible

\begin{figure}[h!]
\includegraphics[width=1.0\linewidth]{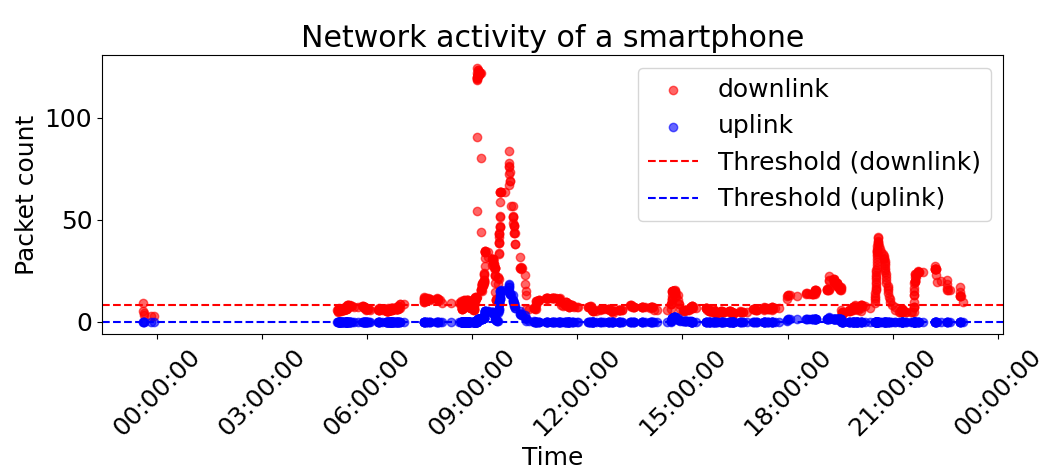}
\caption{Network traffic thresholds of the interactive smartphone (a4:45).}
\label{fig:phone_traffic}
\end{figure}

The \textit{interactive} multimedia devices provide highly legible state changes. As shown in Figure~\ref{fig:phone_traffic}, the smartphone's traffic volume creates distinct thresholds: it remains off until 05:00, idles with minimal background syncs until 09:00, and then transitions into a clear, continuous active state for 1.5 hours. Similar, easily identifiable spikes were observed for the laptop and Smart TV (e.g., heavy downlink traffic spikes correlating with video streaming or boot-up sequences). 

By applying these volume thresholds across the interactive devices, the adversary successfully maps the temporal rhythms of the resident's daily life. They now know \textit{what} devices are in the home, and \textit{when} the resident is actively using them.

\subsection{Device Localization}
With device identities and traffic profiles established, the attacker now aims to understand where these devices are located for a spatial grouping of devices into room-like regions. 

We focus on the stationary devices as they can be used as anchors for room inference. To make this experiment more realistic with the obstruction by walls, we ignore all devices inside the living room. This leaves the five identified smart devices Tuya bulb (d8:f1), Shelly Plus HT (08:b6), Tapo bulb E225 (6c:5a), Tapo smart plug P100 (54:af), Shelly Motion 2 (8c:f6), and the laptop (9c:fc).

\begin{figure}[h!]
\includegraphics[width=\linewidth]{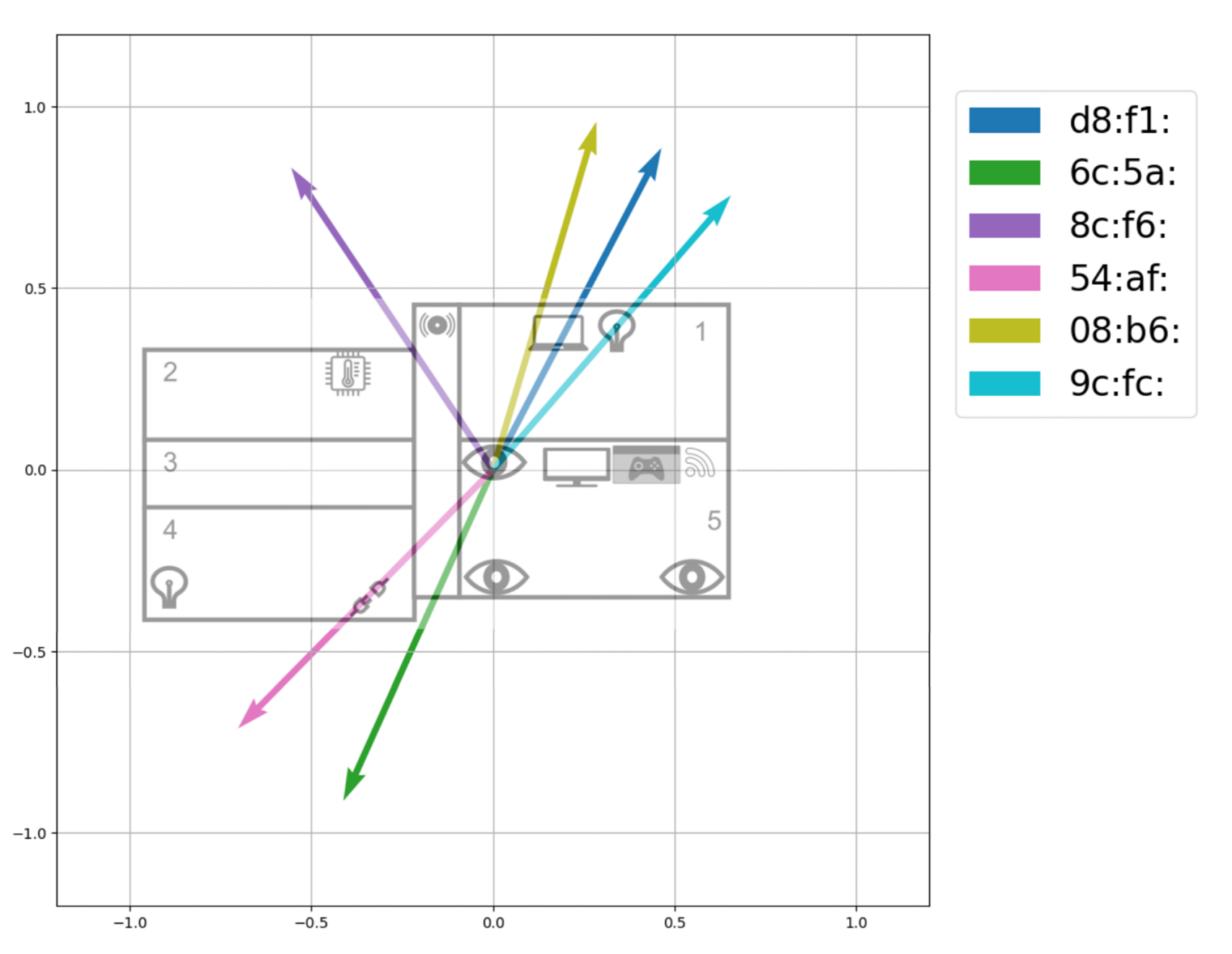}
\caption{Normalized direction vectors of stationary devices based on relative RSSI. The floor plan is shown as ground truth for the reader; the attacker only observes the vectors.}
\label{fig:normalized_vectors}
\end{figure}

For these devices, we construct RSSI ``fingerprints'' by comparing the relative signal strengths received across the three spatially distributed Raspberry Pis. We interpret these fingerprints as 2D direction vectors pointing from the sniffing room towards each device. Figure~\ref{fig:normalized_vectors} visualizes the resulting normalized direction vectors.

The vectors point toward the true locations for four of the six stationary devices. Shelly Plus HT (08:b6, gold) and Tapo bulb E225 (6c:5a, green) are misaligned, likely due to errors caused by wall density and electrical installations between them and the sniffers.

The attacker can still use these vectors to define 2-3 distinct sectors of the house. For example, the vectors for the Tuya bulb (d8:f1, blue) and the laptop (9c:fc, cyan) point into the same area, the attacker could label this as "office" area. Another area can be infered by the positions of the Tapo bulb E225 (6c:5a, green) and Tapo smart plug (54:af, pink).

% Mobiles Tracking
Finally, the same directional mapping can be applied to the resident’s smartphone. Over the course of a day, tracking the phone's changing direction vector allows the attacker to observe movement between these inferred regions. Some of these movement paths never intersect in the angular plots, strongly suggesting the physical presence of walls separating the inferred rooms.

% \begin{figure*}[h!]
% \includegraphics[width=\textwidth]{figures/phone_movement.png}
% \caption{Movement of the inhabitant’s smartphone through the day, represented by changing RSSI-based location estimates relative to the sniffers.}
% \label{fig:smartphone_movement}
% \end{figure*}

\subsection{Human Activity Recognition}
In the final stage of the pipeline, the attacker synthesizes the device inventory, temporal state changes, and spatial mapping to reconstruct the resident's routines.

% \paragraph{External Profiling via Probe Requests}
% Before analyzing indoor habits, the attacker extracts external context from the smartphone's probe requests. Captured probes for specific networks (e.g., \textit{eduroam} and various public hotspots) immediately suggest the resident is affiliated with a university and reveal frequented locations like cafes or shops. This external intelligence helps the attacker build a demographic baseline of the victim before analyzing their home life.

\paragraph{Weekday}
Inspecting a randomly selected weekday, the attacker combines device states and localization to infer specific indoor activities.
\begin{itemize}
    \item \textbf{Work from home:} Sustained laptop traffic originating from the inferred "work area," correlated with the smartphone remaining stationary in that same zone.
    \item \textbf{Breaks and Transitions:} Brief pauses in laptop activity coinciding with smartphone movement to the kitchen or bathroom area.
    \item \textbf{Evening Leisure:} A distinct shift occurs in the late afternoon. Laptop traffic ceases, and the Smart TV and Gaming Console become the dominant traffic sources in the living area, indicating the end of the workday.
\end{itemize}
This synthesis allows the attacker to confidently deduce the resident's daily rhythms, including wake-up times, working hours, and sleep cycles.

\paragraph{Weekly Schedule}

\begin{figure}[h!]
\includegraphics[width=\linewidth]{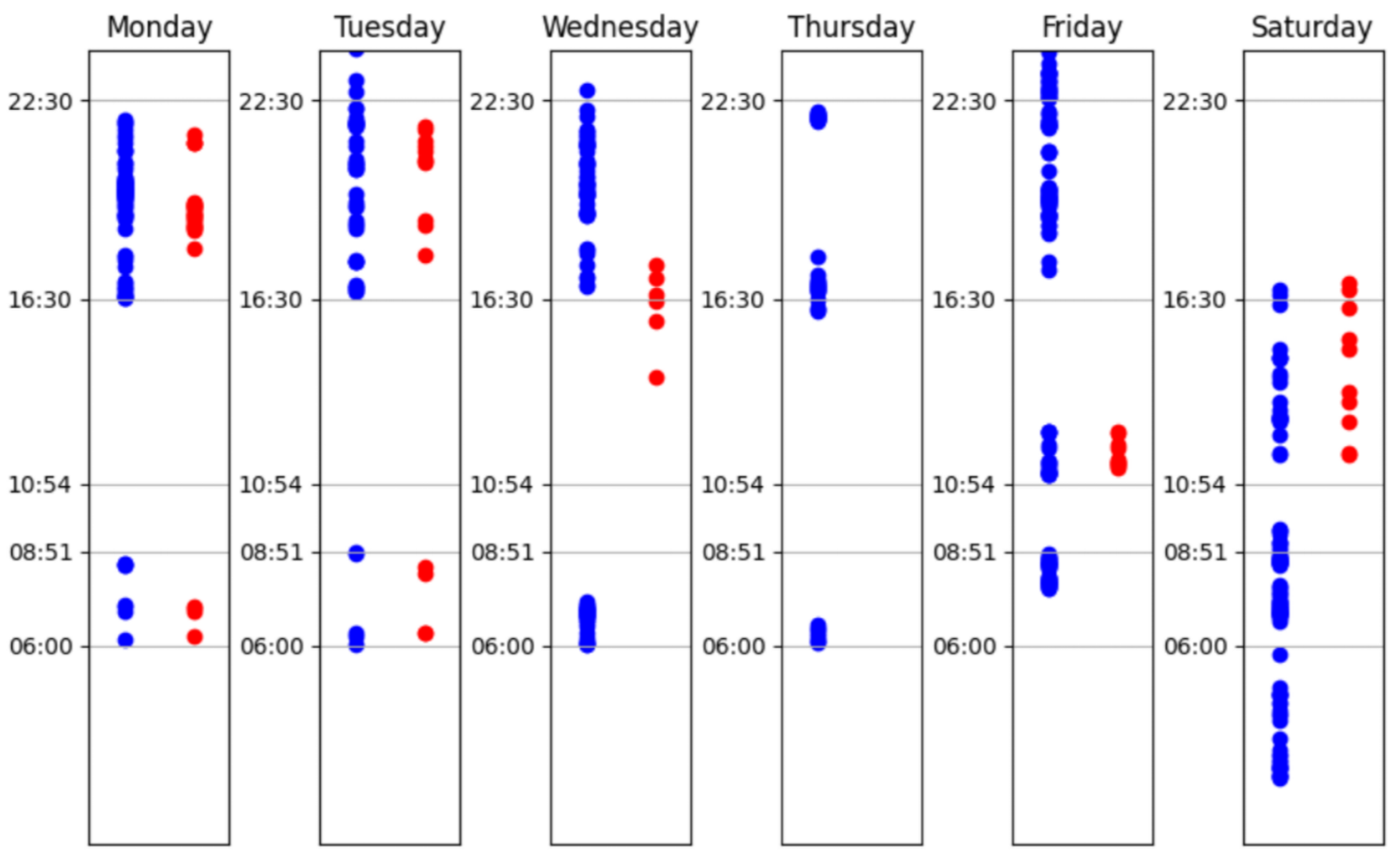}
\caption{Weekly schedule over one week, showing network activity for the smartphone (blue) and laptop (red).}
\label{fig:weekly_schedule}
\end{figure}

By analyzing the presence and absence of interactive devices over a longer time frame (Figure~\ref{fig:weekly_schedule}), the attacker establishes macro-level routines. Rather than guessing specific external activities, the adversary reliably detects recurring periods of absence. For example, consistent, simultaneous drops in smartphone and laptop traffic on specific weekday mornings indicate absence, allowing the attacker to map the resident's standard out-of-home schedule.

% \begin{figure*}[h!]
%     \centering
%     \includegraphics[width=\linewidth]{figures/har_nachmittag.png}
%     \caption{Network activity during a representative afternoon and evening. Red/blue dots represent sent/received 802.11 packets.}
%     \label{fig:har_nachmittag}
% \end{figure*}

\paragraph{Guest Visit}
During the monitoring period, the inhabitant hosted two guests overnight. The attacker immediately detected this event when two unrecognized, interactive devices (smartphones) joined the network.

\begin{figure}[h]
    \centering
    \includegraphics[width=1.0\linewidth]{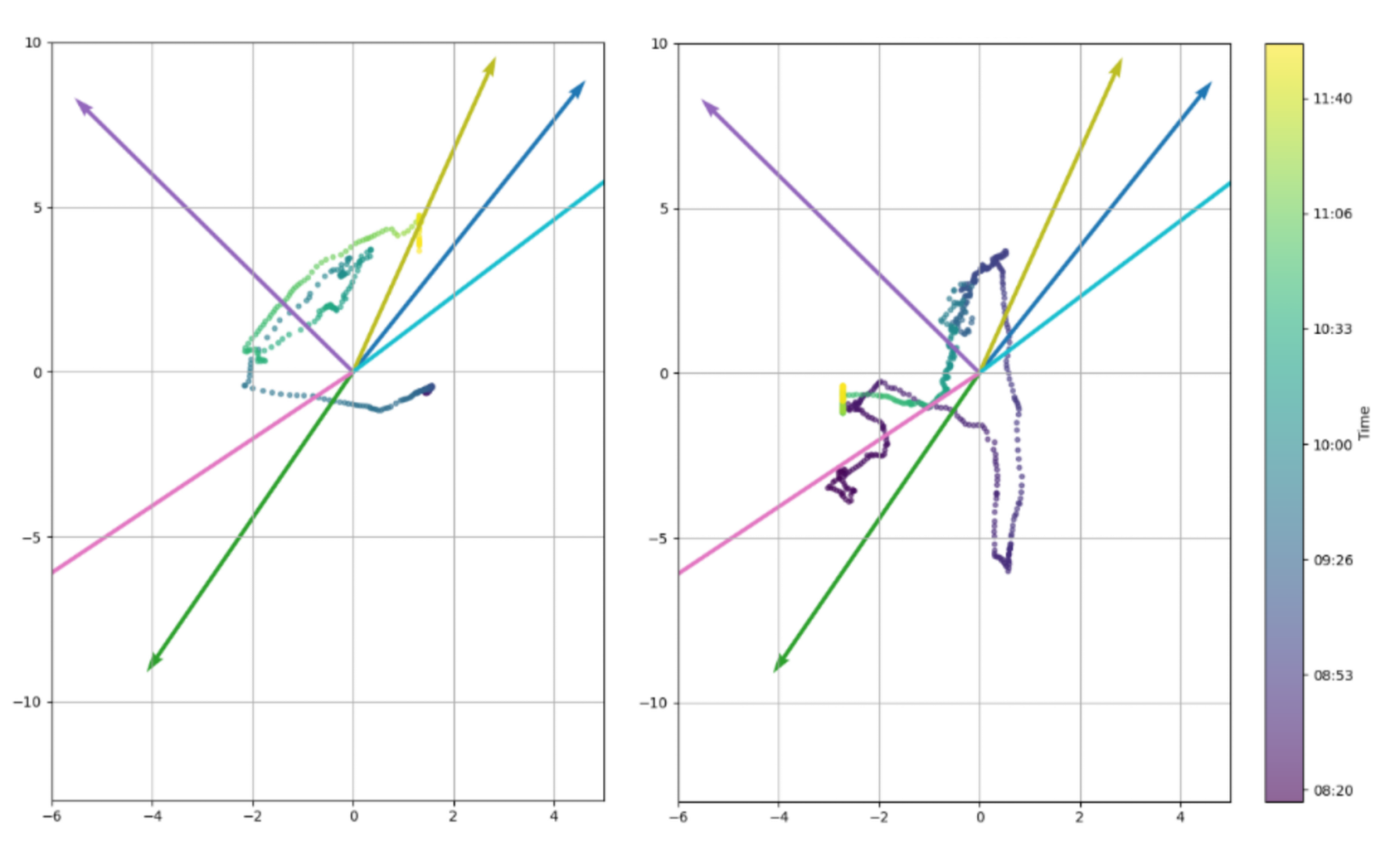}
    \caption{Location estimates of Guest 1 (left) and the inhabitant (right) during the evening.}
    \label{fig:guest_phones}
\end{figure}

By tracking these new devices alongside the baseline inventory, the attacker extracted a detailed timeline of the visit (Table~\ref{tab:guest_visit}). Observing the network traffic of the guest smartphones overnight allowed the attacker to infer the sleep cycles of the visitors. While the resident's and Guest 2's smartphones exhibited flat, minimal background traffic indicative of sleep between 02:00 and 08:00, Guest 1 generated traffic spikes at 04:00, indicating nighttime waking.

\begin{table}[h]
\centering
\caption{Inferred Guest Visit Timeline}
\label{tab:guest_visit}
\begin{tabular}{p{2.5cm} p{5.5cm}}
\toprule
\textbf{Time} & \textbf{Identified Activity} \\
\midrule
21:00 & Guest 1 arrives (new MAC appears) \\
22:00 & Guest 2 arrives (new MAC appears)\\
01:30 -- 02:00 & Transition to sleep (interactive traffic drops)\\
04:00 & Guest 1 wakes up (smartphone active) \\
08:00 -- 09:00 & Inhabitant and Guest 2 wake up \\
$\approx$ 11:45 & Guests leave (MACs disappear)\\
\bottomrule
\end{tabular}
\end{table}

\section{Discussion}
\label{sec:discussion}

Our study demonstrates that a casual attacker can replicate the core privacy-invasive findings of prior work without relying on complex machine learning, curated training data, or reference devices. By applying basic heuristics to unencrypted metadata, we successfully extracted detailed daily routines and sleep cycles. This proves that the technical bar for a ``nosy neighbor'' attack is low. Ultimately, our goal is to inform and warn users about how profoundly vulnerable their private networks are to simple, low-resource eavesdropping.

\paragraph{The Role of the Smartphone vs. Smart Devices}
Unlike most smart home privacy studies, our methodology goes beyond IoT appliances to analyze all wireless devices in the network, reflecting the behavior of a realistic adversary. Consequently, the resident's smartphone proved to be the most revealing source of information. While the smartphone indicates general user activity and movement, it is the smart home appliances that provide the necessary spatial anchors and semantic meaning. The specific type of a smart device immediately reveals what an area is used for, allowing the attacker to effectively profile the physical layout of the home. Furthermore, as other studies have shown, the traffic of these smart devices can be used directly to infer highly specific human activities. Finally, identifying sensitive smart home equipment, such as security cameras or microphones—allows an attacker to build an even more invasive profile of the house, potentially exposing security systems or hardware vulnerabilities. Thus, smart devices remain a critical component for an attacker aiming to comprehensively profile the home.

\paragraph{Limitations of Simplistic Methods}
Our simplified approach does have limitations, particularly regarding device state recognition. While we easily inferred the states of interactive multimedia devices (like the TV and laptop) using simple traffic volume thresholds, this method failed to determine the operational states of autonomous smart appliances (e.g., the Tuya bulb or Shelly sensors). However, related work has proven that state inference for these devices is feasible using packet-level signatures. Therefore, this failure is strictly a limitation of our method, not a lack of underlying vulnerability. Finally, while our setup was realistically obstructed by an interior wall with electrical installations, a true cross-apartment attack might face heavier signal attenuation from walls or insulation.

\section{Conclusion}
\label{sec:conclusion}

% In this work, we showed that an intruder with a simple setup and without physical access could effectively monitor the inhabitant of an adjacent apartment. Furthermore, device were easily distinguishable. The interference from walls has been demonstrated and the findings match the (much more sophisticated) work of \cite{zou2023iotbeholder}.

% \begin{itemize}
%     \item gezeigt, dass Tagesablauf mittels Pipeline gut nachgestellt werden kann
%     \item gezeigt, dass Geräte leicht unterscheidbar sind
%     \item IoTBeholder nachgestellt
%     \item erwartbare Einschränkungen durch Wände, Equipment
%     \item ähnliche Ergebnisse -> Reproduktion geglückt
% \end{itemize}

Our study has revealed that privacy risks in smart-home environments are not limited to highly capable attackers with specialized tools, training data, or machine learning expertise. Even a low-resourced adversary with inexpensive off-the-shelf hardware and simple analysis methods can infer sensitive information about residents’ devices, activities, movements, and daily routines. By demonstrating these attacks in a realistic apartment-building setting over an extended period, we show that the threat is both practical and relevant to ordinary households. This emphasizes the need for privacy mechanisms that also protect against close-by, passive, and persistent eavesdroppers of wireless network traffic.

\printbibliography

@article{Li2025WiFingerFN,
  title={Wi{F}inger: Fingerprinting Noisy Io{T} Event Traffic Using Packet-level Sequence Matching},
  author={Ronghua Li and Shinan Liu and Haibo Hu and Qingqing Ye and Nick Feamster},
  journal={ArXiv},
  year={2025},
  volume={abs/2508.03151}
}

@article{zou2023iotbeholder,
author = {Zou, Qingsong and Li, Qing and Li, Ruoyu and Huang, Yucheng and Tyson, Gareth and Xiao, Jingyu and Jiang, Yong},
title = {Io{TB}eholder: A Privacy Snooping Attack on User Habitual Behaviors from Smart Home {W}i-{F}i Traffic},
year = {2023},
issue_date = {March 2023},
publisher = {ACM},
% address = {New York, NY, USA},
volume = {7},
number = {1},
% doi = {10.1145/3580890},
journal = {Proc. ACM Interact. Mob. Wearable Ubiquitous Technol.},
month = mar,
articleno = {43},
numpages = {26},
keywords = {Internet-of-Things, Privacy, Smart Home}
}

@article{apthorpe2017spying,
  title={Spying on the smart home: Privacy attacks and defenses on encrypted {I}o{T} traffic},
  author={Apthorpe, Noah and Reisman, Dillon and Sundaresan, Srikanth and Narayanan, Arvind and Feamster, Nick},
  journal={ArXiv},
  year={2017}
}

@inproceedings{acar2020peekaboo,
author = {Acar, Abbas and Fereidooni, Hossein and Abera, Tigist and Sikder, Amit Kumar and Miettinen, Markus and Aksu, Hidayet and Conti, Mauro and Sadeghi, Ahmad-Reza and Uluagac, Selcuk},
title = {Peek-a-{B}oo: i see your smart home activities, even encrypted!},
year = {2020},
%isbn = {9781450380065},
publisher = {ACM},
%address = {New York, NY, USA},
%url = {https://doi.org/10.1145/3395351.3399421},
%doi = {10.1145/3395351.3399421},
booktitle = {Proceedings of the 13th ACM Conference on Security and Privacy in Wireless and Mobile Networks},
pages = {207–218},
numpages = {12},
keywords = {BLE, ZigBee, network traffic, privacy, smart-home, wifi},
%location = {Linz, Austria},
%series = {WiSec '20}
}

@article{trimananda2019pingpong,
  title={Ping{P}ong: Packet-level signatures for smart home device events},
  author={Trimananda, Rahmadi and Varmarken, Janus and Markopoulou, Athina and Demsky, Brian},
  journal={Proc. NDSS},
  year={2020},
  doi={https://dx.doi.org/10.14722/ndss.2020.24097}
}

@article{apthorpe2016poster,
  title={Poster: A smart home is no castle: Privacy vulnerabilities of encrypted iot traffic},
  author={Apthorpe, Noah and Reisman, Dillon and Feamster, Nick},
  journal={Proc. NDSS},
  year={2016},
  url={https://api.semanticscholar.org/CorpusID:15706672}
}

@article{Wang2020FingerprintingEV,
  title={Fingerprinting encrypted voice traffic on smart speakers with deep learning},
  author={Chenggang Wang and Sean Kennedy and Haipeng Li and King Hudson and Gowtham Atluri and Xuetao Wei and Wenhai Sun and Boyang Wang},
  journal={Proceedings of the 13th ACM Conference on Security and Privacy in Wireless and Mobile Networks},
  year={2020}
  %url={https://api.semanticscholar.org/CorpusID:218720015}
}

@article{Wang2024ISS,
  title={I Still See You: Why Existing {I}o{T} Traffic Reshaping Fails},
  author={Su Wang and Keyang Yu and Qi Li and Dong Chen},
  journal={ArXiv},
  year={2024},
  volume={abs/2406.10358},
  url={https://api.semanticscholar.org/CorpusID:270560248}
}

@article{mane2026rssi_distance_ai,
author = {Mane, Sarika and Kulkarni, Makarand and Gupta, Sudha},
title = {{RSSI}-Based Indoor Distance Estimation in {W}i-{F}i {I}o{T} Application Using AI Approaches},
journal = {International Journal of Communication Systems},
volume = {38},
number = {5},
year = {2025}
}

@INPROCEEDINGS{Iglesias2012IndoorPL,
  author={Pérez Iglesias, Héctor José and Barral, Valentín and Escudero, Carlos J.},
  booktitle={2012 19th International Conference on Systems, Signals and Image Processing (IWSSIP)}, 
  title={Indoor person localization system through {RSSI} {B}luetooth fingerprinting}, 
  year={2012},
  volume={},
  number={},
  pages={40-43},
  doi={}
}

@INPROCEEDINGS{bahl2000RADAR,
  author={Bahl, P. and Padmanabhan, V.N.},
  booktitle={Proceedings IEEE INFOCOM 2000},  
  %Conference on Computer Communications. Nineteenth Annual Joint Conference of the IEEE Computer and Communications Societies
  title={RADAR: an in-building RF-based user location and tracking system}, 
  year={2000},
  volume={2},
  number={},
  pages={775-784 vol.2},
  doi={10.1109/INFCOM.2000.832252}
}

@misc{wireshark,
    title={Wireshark},
    author={Wireshark Foundation},
    url={https://www.wireshark.org/},
    note={Accessed: 2026-02-05}
}

@misc{tcpdump,
  title        = {tcpdump - Packet Analyzer},
  author       = {{tcpdump.org}},
  howpublished = {\url{https://www.tcpdump.org/}},
  note         = {Accessed: 2025-03-26}
  %year         = {2025}
}

@misc{bluepy,
    title={bluepy: Python interface to Bluetooth {LE} on Linux},
    author={Ian Harvey},
    url={https://github.com/IanHarvey/bluepy},
    note={Accessed: 2026-02-05}
}

@misc{macvendors,
  title        = "{MA:CV:en:do:rs}",
  howpublished = {\url{https://macvendors.com/}},
  note         = {Accessed: 2026-02-23}
}

@misc{macaddressio,
  title        = "{MAC Address Vendor Lookup}",
  howpublished = {\url{https://macaddress.io/}},
  note         = {Accessed: 2026-02-23}
}

@mastersthesis{bartosz2025,
    author={Bartosz Wojciech Burgiel},
    title={Noisy Networks, Nosy Neighbors: Inferring Privacy Invasive Information from Encrypted Wireless Traffic},

    type={Bachelor's Thesis},
    note={ArXiv},
    year={2025}
}

\end{document}